\documentstyle[epsf]{article}
\begin{document}

\begin{center}
{\Large\bf
Temperature dependence of the ``0.7'' 2e$^2$/h 
quasi plateau in strongly confined quantum point contacts}\\[2mm]
{\large A.~Kristensen, P.E.~Lindelof, J.~Bo Jensen, M.~Zaffalon, 
J.~Hollingbery, S.W.~Pedersen,\\[1mm] 
J.~Nyg\aa rd, H.~Bruus, S.M.~Reimann, and C.B.~S\o rensen}\\
{\em Niels Bohr Institute, \O rsted Laboratory, 
Universitetsparken 5, DK-2100 Copenhagen \O, Denmark}\\[2mm]
{\large M.~Michel and A.~Forchel}\\
{\em Technische Physik, Universit\"at W\"urzburg,
Am Hubland, D-97074 W\"urzburg, Germany}\\
\vspace*{2mm} 
(To appear in Physica B, 1998)
\end{center}

\centerline{\bf Abstract}
\noindent
We present new results of the ``0.7''$(2 e^{2}/h)$ structure 
or quasi plateau in some of the most 
strongly confined point contacts so far reported. This strong
confinement is obtained by a combination of shallow etching and metal
gate deposition on modulation doped GaAs/GaAlAs heterostructures.
The resulting subband separations 
are up to 20 meV, and as a consequence the quantized conductance can
be followed at temperatures up to 30 K, an order of magnitude
higher than in conventional split gate devices. We observe pronounced
quasi plateaus at several of the lowest conductance steps all the
way from their formation around 1 K to 30 K, where the entire 
conductance quantization is smeared out thermally. We study the
deviation of the conductance from ideal integer quantization as a
function of temperature, and we find an activated behavior,
$\exp(-T_a/T)$, with a density dependent activation temperature $T_a$
of the order of 2 K. We analyze our results in terms of a simple
theoretical model involving scattering against plasmons in the
constriction.\\

\centerline{\bf I. Introduction}
\noindent
The quantized conduction through a narrow point contact is one of the
key effects in mesoscopic physics believed to demonstrate e.g.\ the
validity of the single particle Fermi liquid picture in terms of the
Landauer-B\"{u}ttiker formalism, a central formalism in the field. 
However, the electron system in the narrow constriction forms a quasi 
one-dimensional electron liquid, and such systems have long ago been
predicted to exhibit significant deviations from the ordinary Fermi
liquid behavior. Thus the quantum point contact (QPC) remains an
important testing ground for the description of mesoscopic phenomena. Indeed
recently, significant deviations from the Landauer-B\"{u}ttiker theory
have in fact been observed in quantum point contacts in the
temperature dependence of the conductance quantization
\cite{Tarucha95,Yacoby96} and as a so called ``0.7'' structure
or quasi plateau,  appearing around $0.7$ times the conductance quantum $2 e^{2}/h$
\cite{Thomas95}. 
Invoking
a Luttinger liquid approach \cite{Kane92} the deviations have been
discussed in terms of interaction effects
\cite{Kawabata96,Shimizu96,Oreg96} and spin polarization of the
one-dimensional electron liquid \cite{Schmeltzer97}. 
However, firm conclusions have 
been difficult to obtain partly due to the narrow temperature range
(0.1 K - 4 K) in which the effect can be studied in conventional split
gate quantum point contacts, where relatively close lying
one-dimensional subbands are formed. One major point in this work is
the fabrication of QPCs with large subband spacings which allow a more
detailed study of the temperature dependence of the deviations from
the standard single particle picture.

\vspace*{2mm} 
\centerline{\bf II. The device}
\noindent
Our quantum point contacts were fabricated on conventional high electron
mobility transistor (HEMT) structures.
By Molecular Beam Epitaxy (MBE)  the following layer sequence was
grown: 1 $\mu$m ${\rm GaAs}$
buffer, $20 \, {\rm nm}$ ${\rm Ga_{0.7} Al_{0.3} As}$ spacer,
$40 \, {\rm nm}$ ${\rm Ga_{0.7} Al_{0.3} As}$ barrier layer with a Si
concentration of $2 \times 10^{18} \; {\rm cm^{-3}}$ and a 10~nm
GaAs cap layer. 
The two-dimensional electron gas forms in the GaAs at the
${\rm GaAs/Ga_{0.7}Al_{0.3}As}$ interface, buried $90 \, {\rm nm}$ below the
surface of the heterostructure.
At 4.2 K the carrier density was $2 \times 10^{15} \, {\rm m^{-2}}$
and the mobility $104 \, {\rm m^2/Vs}$, measured in the dark.
The sample was processed with a $20 \times 100 \, {\rm \mu m^{2}}$ mesa,
etched $100 \, {\rm nm}$, and AuGeNi ohmic contacts to the 2DEG were formed
by conventional UV-litho\-gra\-phy and lift-off. The narrow QPC
constriction was defined by shallow etching on the mesa. In order to
perform the shallow etch with well-controlled edges, a $13 \, {\rm
nm}$ thick Al  mask was formed by e-beam lithography and lift-off,
using a field emission scanning electron microscope at an acceleration
voltage of $2.5 \, {\rm kV}$. The sample was then shallow etched $55 -
60 \, {\rm nm}$ in ${\rm H_{2}O : H_{2} O_{2} : NH_{4}OH}$
$(75:0.5:0.5)$ at an etch rate of $4 \, {\rm nm/sec}$. The etch-mask
was removed, and a $80 \, {\rm nm}$ thick, $5 \, {\rm \mu m}$ wide Al
gate electrode was deposited by UV-lithography and lift-off, covering
the constriction and the surrounding 2DEG. The shape $y(x)$ of the
etched constriction is parabolic, $y(x) = \pm ( d/2 + a \,
x^{2})$. Hence the geometry of the QPC is characterized by two
numbers, the width of the constriction, $d$, and the curvature, $a$.
We have studied values of $d$ between 50 nm and 400 nm and of
$a$ between 0.00125~nm$^{-1}$ and 0.005~nm$^{-1}$. For channel widths
$d < 300$~nm the channel is pinched-off at zero gate voltage, and a 
positive gate voltage is applied to open it. The conductance of
the QPC was measured by conventional lock-in technique using an ac
current bias of 1~nA at a frequency of 17~Hz. The gate leakage current
was also monitored during low temperature measurements, and it was
found to be negligible (less than our measurement resolution of 10~pA)
for $V_{G} < 1$~V. All samples showed
well-defined conductance quantization  well above 4 K. The most
regular conductance steps at the highest temperatures were seen for
the narrowest and longest constrictions $d = 50$ nm, 
$a = 0.00125$ nm$^{-1}$.

We estimate subband separations in our QPCs up to $20$~meV 
by comparing the
measured temperature dependence to the theoretical model for
conductance quantisation \cite{Fertig87,Buttiker90}, and by applying a
finite dc source drain voltage \cite{Thomas95}. 
These are the highest
values so far reported for lateral QPCs in ${\rm GaAlAs/GaAs}$
heterostructures \cite{Ismail90}. They are much higher than
the corresponding ones found in split gate devices
\cite{Taboryski95,Thomas95} where subband energy spacings of $\simeq 2
\; {\rm meV}$ are found, and where the conductance quantisation
disappears at about $4 \; {\rm K}$.

\vspace*{2mm} 
\centerline{\bf III. Experimental results}
\noindent
In this paper we focus on the results of a particular QPC with an
etched center width $d = 205$~nm and wall curvature $a =
0.00125$~nm$^{-1}$. The discussed temperature dependence is
reproduced in four other samples investigated. 
Even though the quantization of conductance in the sample is
visible up to 30 K, we present in Fig.~1 only
measurements from 0.3 K to 8 K. In this temperature regime the
deviations from 
\begin{figure}[h]
\makebox[8cm]{\epsfysize=70mm\epsfbox{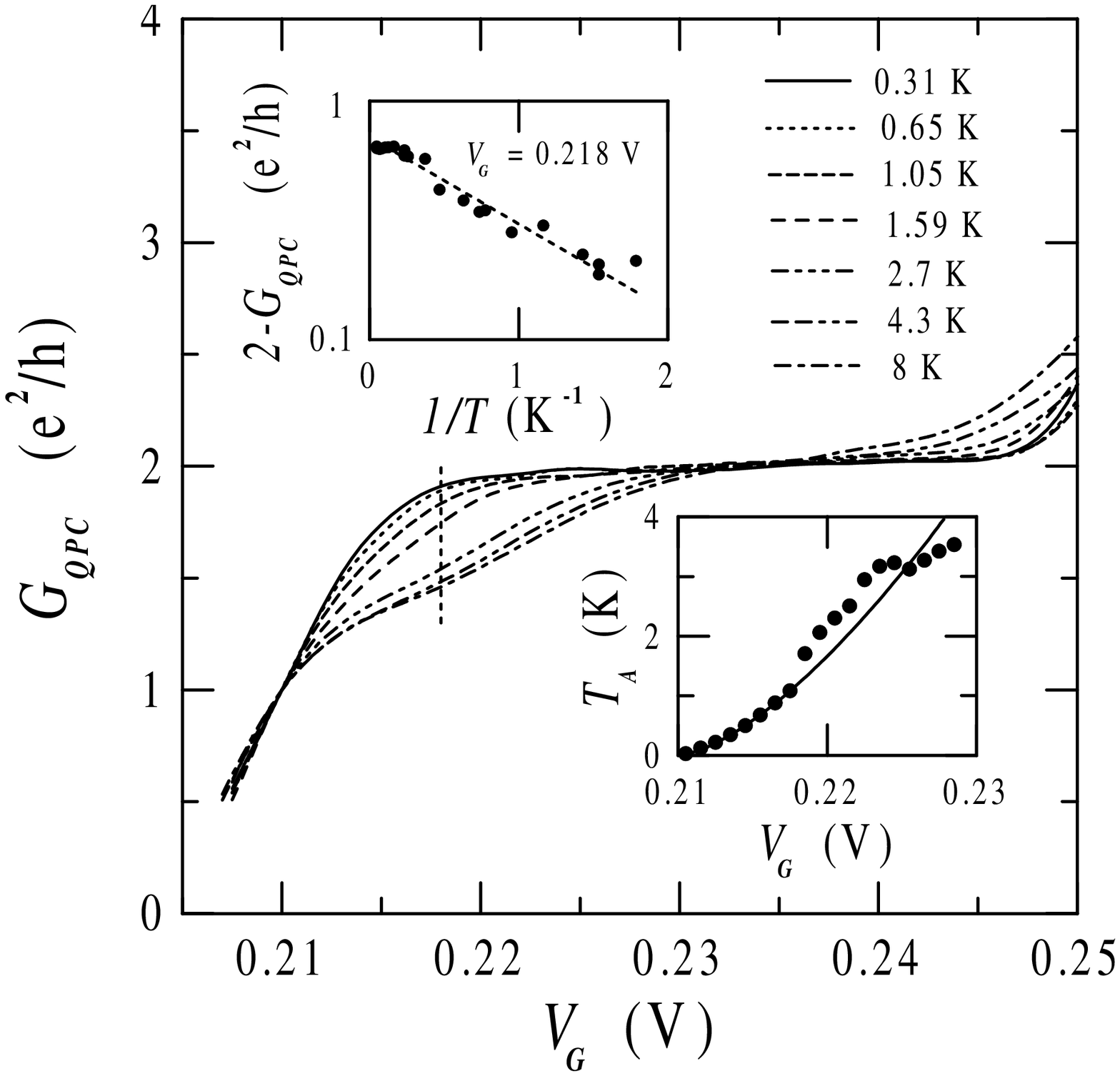}} \hfill
\makebox[8cm]{\epsfysize=70mm\epsfbox{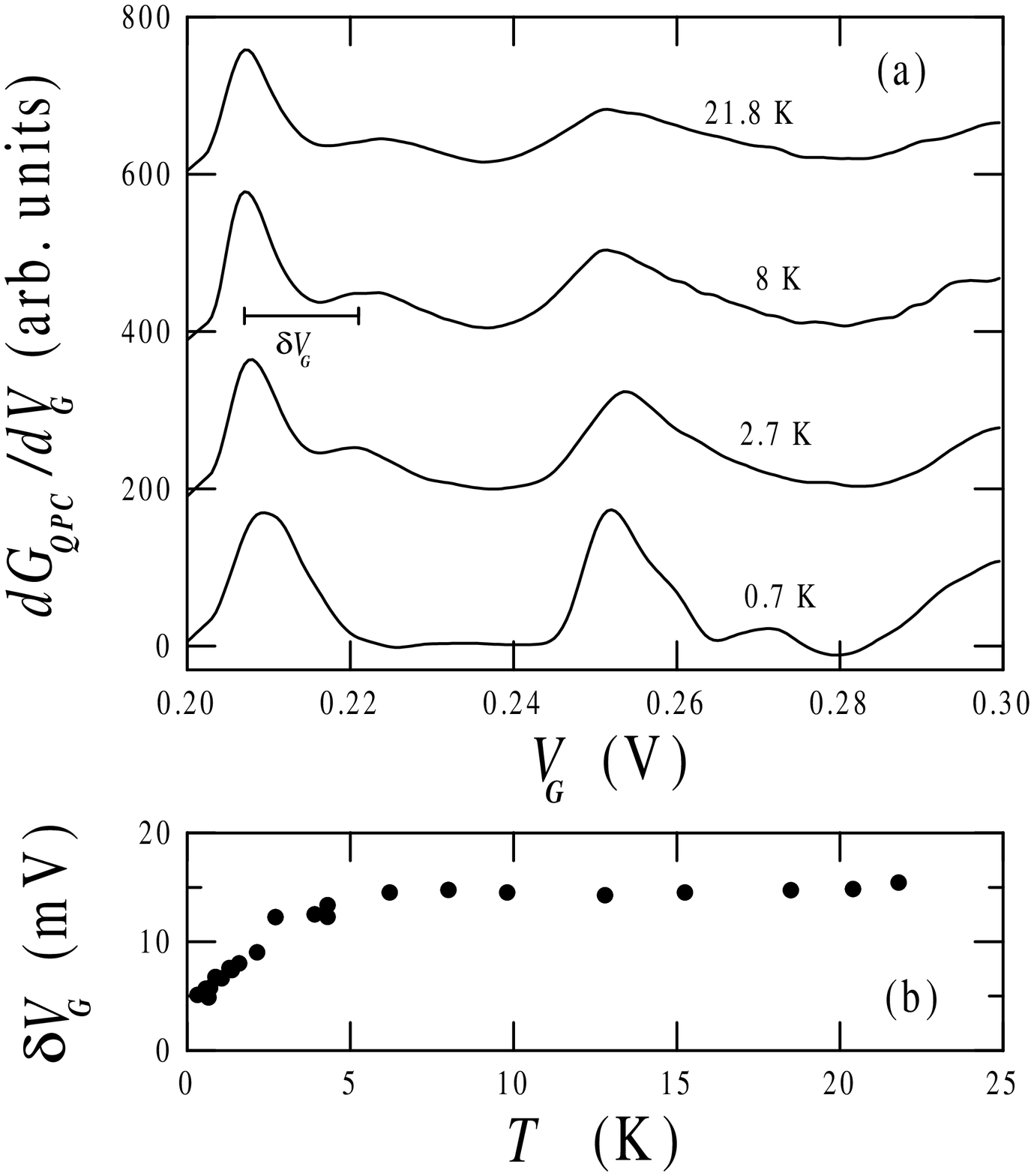}}
\begin{minipage}[t]{8cm} \noindent 
FIG.\ 1.
Quantum point contact conductance $G_{QPC}$ versus gate voltage $V_G$
at the first conductance plateau measured at different
temperatures. The QPC has a width $d = 205$~nm and a curvature
$a = 0.00125$~nm$^{-1}$ (see text). Upper insert: Arrhenius plot of
the deviation from ideal quantized conductance, $(2-G_{QPC})$ 
versus $1/T$ at gate voltage $V_{G}= 0.218$~V (indicated
by the vertical dashed line on the $G_{QPC}$ -- $V_G$ plot). The
slope of dashed line on the Arrhenius plot gives an activation
temperature $T_a$. Lower insert: $T_a$ versus $V_G$ for the first half
of the first conductance plateau. The full line is the result of the
model calculation of section IV.
\end{minipage} \hfill
\protect\begin{minipage}[t]{8cm} \noindent
FIG.\ 2.
(a) Transconductance $dG_{QPC}/dV_{G}$ evaluated from the same data set
as in Fig.~1. The ``0.7''~(2$ne^2/h$) structure
emerges around 1~K as a satellite to the first transconductance peak
$(n=1)$, and as shoulders on the peaks at higher $n$. The
``0.7''-satellite can be followed to the highest temperatures where
conductance quantization is still visible.
(b) Separation $\delta V_{G}$ between the $n=1$ main peak
and the ``0.7''-satellite as a function of temperature. 
\end{minipage}
\end{figure}
Landauer-B\"uttiker theory is most clearly seen
without having too much thermal smearing present. The main observation
is that the deviation from perfect quantization seems to follow an
activated behavior. This is demonstrated in the upper insert of the
figure: a semi-logarithmic plot of  the deviation $2-G_{QPC}$ versus
$1/T$ at a gate voltage of $V_{G} = 0.218$~V. In the temperature
region from 1~K to 8~K the data exhibit an 
activated behavior,
from which we extract an activation temperature $T_{A}$. 
However, some caution is needed: our data only allow us to plot the
activation behavior over one single order of magnitude.
The lower insert of Fig.~1 shows the extracted activation temperature
as a function of gate voltage across the lower part of the first
conductance plateau. We find that it rises from 0 K to 4 K as the gate
voltage is swept through the conductance step, i.e.\ as the electron
density grows. This feature is discussed in section IV. 

In Fig.~2a we show the transconductance
$dG_{QPC}/dV_{G}$ measured at different temperatures. The first two
conductance plateaus are seen as minima in transconductance at $V_{G}
\sim 0.235$ and $0.28$~V,  separated by transconductance peaks. At
temperatures above 1~K the ``0.7'' quasi plateau appears  at
$V_{G} \sim 0.225$~V as a satellite to the first conductance peak, and
as a shoulder to the higher peaks. From this behavior we can rule out
thermal smearing as the mechanism behind the observed
structure. Thermal smearing would broaden the transconductance peaks
and make them more symmetric. 
The particular shape of the transconductance peaks could also simply
be reflecting the energy dependence of the transmission probability
for electrons traversing the constriction. However, such features
would again be smeared out thermally rather than giving rise to
additional structure appearing at high temperature. 
In Fig.~2b we have plotted the distance $\delta
V_{G}$ between the main peak and the satellite peak. The distance
grows at the lowest temperatures, but above 5~K it saturates.

Our experimental results represent a confirmation and extension of the
earlier results by Thomas et al \cite{Thomas96}, who suggest that the
``0.7''-structure may originate from spin-polarization of the
1D-electron gas in zero magnetic field. 
We have also measured the low temperature QPC conductance with a
strong magnetic field, $B \leq 12$~T, applied in-plane with the 2DEG,
and parallel with the current through the constriction. 
Like Thomas et al.\ we observe that the separation $\delta V_{G}$
between the first main and satellite  peak increases linearly with
$B$, and we extract an enhanced g-factor of $g = 0.66$. 

\vspace*{2mm} 
\centerline{\bf IV. A theoretical model}
\noindent
The quasi one-dimensionality of the electron liquid in the QPC
has brought forward the possibility of Luttinger liquid behavior 
\cite{Kane92}. The quasi plateau has been discussed in the
characteristic terms of separated charge and spin degrees of freedom
of one-dimensional strongly coupled systems
\cite{Kawabata96,Shimizu96,Oreg96,Schmeltzer97}. However, explanations
invoking the Luttinger liquid face at least one serious problem. The
channel length of the typical QPC is very short, of the order of ten
times the Fermi wave length of the bulk 2DEG and only a few times the
Fermi wave length or less of the 1D channel itself. In contrast, the 
characteristic correlations of the Luttinger liquid are established
over long ranges. 
We are therefore seeking the explanation of the quasi plateau in more
simple terms. As mentioned, the evolution of the quasi plateau as a
function of temperature rules out single particle effects like thermal
smearing, asymmetrical barrier potential and resonances. It seems
necessary to include electron-electron interaction effects to explain
the phenomena. We propose a model, where the quasi plateau is due to
electron-electron interaction in the form of scattering of electrons
against plasmons in the QPC. 

We calculate the frequency $\omega^{1D}_p$ of the quasi
one-dimensional plasmons of the QPC in the long wave limit of the
random phase approximation. The QPC is embedded in a dielectric of
permittivity $\varepsilon$, and it is compensated by uniform
background of the same density. The resulting dispersion relation
is 

\begin{equation} \label{omega}
\omega^{1D}_p = \sqrt{v_f^2 + 
\frac{\gamma e^2 n^{1D} }{4 \pi \varepsilon m^*}} \; q, 
\end{equation}
where $m^*$ is the effective electron mass and $\gamma \approx 1$ is a
numerical factor. Due to the presense of the 2DEG contacts the 1D
plasmons are size quantized and acquire a finite gap. The 
lowest lying mode is a standing half wave with a wave number of the
order of $q=\pi/L$, where $L$ is the channel length. 
The electron density $n^{1D}$ is
controlled by the gate voltage $V_G$ through 
$n^{1D} \propto C (V_G - V_0)$,
where $C$ is the capacitance and $V_0$ the pinch-off voltage, here
defined by $G(V_0) = e^2/h$. As $V_G$ is increased from $V_0$ the
electron channel width increases from zero, and hence the capacitance
is also an increasing function of $V_G$. Since $n^{1D}$ is
proportional to $\sqrt{\varepsilon_F}$, where $\varepsilon_F$ is
the Fermi energy in the channel, we end up with
$\varepsilon_F = \alpha (V_G - V_0)^3$ to lowest order in
$V_G$. The constants $V_0$ and $\alpha$ are found by combining the
estimates of the 1D sublevel spacings with the two values of $V_G$
where $G(V_G) = e^2/h$ and $G(V_G) = 3e^2/h$, respectively. Finally we
fit the resulting plasmon energy $\hbar \omega^{1D}_p$ to the
activation data of Fig.~1 by adjusting $L$. In the
insert of Fig.~1 we used $L = 0.1\;\mu$m.

The justification of our model is the following. The electron gas
is divided into three regions, two 2DEG contact regions and one 1DEG
QPC region. Scattering against plasmons of the bulk 2DEG is always
present at all temperatures, and it is taken into account through the
background resistance, which as usual is subtracted to obtain the correct
values of the quantization plateaus. At temperatures well below $\hbar
\omega^{1D}_p$ no further resistance is introduced by the 1D plasmons
in the QPC. As temperature increases scattering against
1D plasmons becomes possible and results in an additional resistance
leading to deviation from perfect conductance quantization of an
activated form, with the activation energy given by the plasmon
energy. As the gate voltage is increased the density of the 1DEG is
also increased, and the gap in the plasmon spectrum grows, resulting
in an enhanced activation energy. Finally, the plasmons of the 1DEG becomes
more like the bulk 2DEG plasmons, and the additional resistance
disappears into the background resistance. The position of the quasi
plateau is given by the voltage required to have a sufficient
density of the 1DEG to have significant scattering against the
plasmons. Of course, this crude model needs to be extended by taking
into account for example possible spin polarization effects and the
opening of the next conducting channel.

\vspace*{2mm} 
\centerline{\bf V. Concluding remarks}
\noindent
We have fabricated lateral QPCs in GaAs with an extraordinary strong
confinement potential in the constriction raising by an order of
magnitude the maximal temperature at which quantized conductance can
be observed as compared to conventional split-gate devices. The
enlarged temperature range enabled us to observe activated temperature
dependence of the ``0.7'' quasi plateaus, and we measured the
activation temperature to rise from 0 K to 4 K as the gate voltage is
increased through the first half of the first conductance plateau.
We have suggested a model to explain the observed effects based on
scattering against 1D plasmons in the QPC. The model introduces a gap
in the plasmon spectrum in a natural way. This gap is identified 
with the activation temperature, and the model can account for the
gate voltage dependence of that temperature.

\vspace*{2mm} 
\centerline{\bf Acknowledgements}
\noindent
This research is part of the EU IT-LTR program Q-SWITCH (No.\ 20960).
The III-V materials used in this investigation were made at the
III-V NA\-NO\-LAB, operated jointly by the Microelectronics Centre of the
Danish Technical Universtiy and the \O rsted Laboratory, Niels Bohr
Institute fAFG, University of Copenhagen. The research was partly
supported by SNF grants 9502937 and 9601677. S.M.R. thanks the
Studienstiftung des deutschen Volkes and the BASF AG for financial
support. H.B.\ is supported by the Danish Natural Science Research
Council through Ole R\o mer Grant no.\ 9600548.

\end{document}